\documentclass[preprint]{aastex701}

\begin{document}

\title{A Practical Framework for Weight Map Construction in CCD Photometry}

\author[orcid=0000-0002-0409-5719]{Dezi Liu}
\affiliation{South-Western Institute For Astronomy Research, Yunnan University,  Kunming 650500, China}
\affiliation{Key Laboratory of Survey Science of Yunnan Province, Yunnan University, Kunming 650500, China}
\email[show]{adzliu@ynu.edu.cn}

\author[orcid=0009-0006-1010-1325]{Yuan Fang}
\affiliation{South-Western Institute For Astronomy Research, Yunnan University,  Kunming 650500, China}
\affiliation{Key Laboratory of Survey Science of Yunnan Province, Yunnan University, Kunming 650500, China}
\email[show]{fangyuan@ynu.edu.cn}

\begin{abstract}

This note presents a practical formulation for constructing weight maps in CCD photometry, accounting for various noise contributions under the assumption of statistical independence. We provide complete equations for both electron-count and ADU-based imaging data, along with simplified expressions for different observational conditions.

\end{abstract}



\section{introduction} \label{sec:intro}
Weight maps are essential for source detection, flux uncertainty estimation, and image co-addition in CCD photometry. By definition, the weight map of an image represents the inverse of the pixel-level noise variance. The dominant noise contributions include Poisson noises from the background and astronomical objects, as well as detector read noise. However, studies in the literature lack detailed derivations of weight maps \citep[e.g.,][]{2005AN....326..432E,2017PASP..129k4502B}. This motivates the presentation of a practical formulation in this note. Table~\ref{tab:param} lists the symbols with their definitions and units used in our analysis.

Mathematically, the two-dimensional raw image $R$ can be expressed as
\begin{equation}\label{eq:raw_image}
    R = (T+S)\times F + D + B.
\end{equation}
We have ignored other effects, such as nonlinearity, fringing, and crosstalk. These additional multiplicative and additive terms can be easily incorporated into weight map construction by following a similar approach outlined in this note.

To obtain the science image $T$, we generally need to first construct master calibration frames $B_{m}$, $D_{m}$, and $F_{m}$. After their calibrations, we then model the background and subtract it from the calibrated image. Following the standard image processing procedures, $T$ is obtained through  
\begin{equation}\label{eq:sci_image}
    T = \frac{R- B_{m} - D_{m}}{F_{m}} - S_{m}.
\end{equation}
The noises in the different terms on the right-hand side of equation~(\ref{eq:sci_image}) will propagate into $T$, so we can use this equation as a starting point to formulate the variance map of $T$, denoted as $V(T)$. Once we obtain $V(T)$, the weight map $W(T)$ can be directly derived by $W(T) = 1/V(T)$. 

\begin{table}[]
    \centering
    \begin{tabular}{|c|l|c|}
    \hline
    Symbol & Definition & Unit \\
    \hline
    $R$    &  raw image & $e^{-}$ \\
    \hline
    $T$    &  science image  &  $e^{-}$ \\
    \hline
    $S/S_{m}$    &  true background/background model & $e^{-}$ \\
    \hline
    $B/B_{m}$ & single bias frame/master bias & $e^{-}$ \\
    \hline
    $D/D_{m}$ & single dark current frame/master dark current & $e^{-}$ \\
    \hline
    $F/F_{m}$ & single normalized flat frame/master flat & - \\
    \hline
    $N_{B}$ & number of single bias frames & - \\
    \hline
    $N_{D}$ & number of single dark current frames & - \\
    \hline
    $N_{F}$ & number of single flat frames & - \\
    \hline
    $\sigma_\mathrm{RD}$ & read noise & $e^{-}$/pixel \\
    \hline
    $g$ & gain & $e^{-}$/ADU \\
    \hline
    \end{tabular}
    \caption{List of symbols with their definitions and units used in this study.}
    \label{tab:param}
\end{table}

\section{methodology} \label{sec:method}
A basic assumption to derive $V(T)$ is that different noises are independent. Therefore, the error propagation equation can be applied. Immediately, $V(T)$ can be expressed as 
\begin{equation}\label{eq:vt1}
    V(T) = \frac{1}{F_{m}^2}\left[V(R)+V(B_{m})+V(D_{m})\right] + \frac{(T+S_{m})^2}{F_{m}^2}V(F_{m}).
\end{equation}
In this equation, we have ignored the uncertainty of $S_{m}$. In what follows, we will quantify the variance of each term.
\begin{itemize}
    \item Master bias $B_{m}$. We typically construct $B_{m}$  by averaging $N_{B}$ individual bias frames, i.e., $B_{m} = \sum_{i=1}^{N_{B}}B_{i}/N_{B}$. However, before averaging, a sigma-clipping algorithm is often applied to reject bad pixels. As a result, the effective number of frames $N_{B,\mathrm{eff}}$ contributing to each pixel in $B_{m}$ is position-dependent and satisfies $N_{B,\mathrm{eff}} \leq N_{B}$. This consideration also applies to the construction of the master dark current and flat. Consequently, the effective number of frames should be represented as a two-dimensional map instead of a single number. Here for simplicity, we retain $N_{B}$ for the following analysis. For each $B_{i}$, the predominant noise is the read noise $\sigma_\mathrm{RD}$, which follows a Gaussian distribution. Therefore, the variance of $B_{m}$ is 
    \begin{equation}\label{eq:vbm}
        V(B_{m}) = \frac{\sigma_\mathrm{RD}^2}{N_{B}}.
    \end{equation}
    
    \item Master dark current $D_{m}$. We also take $N_{D}$ single dark current frames to construct $D_{m}$, i.e., $D_{m}=\sum_{i=1}^{N_{D}}D_{i}/N_{D}$. Note that the raw dark current frames taken directly from the detector contain bias, thus we have to subtract $B_{m}$ individually before averaging. For each $D_{i}$, the noise contributions are from thermal electrons, which follows Poisson statistics, and read noise. The variance of the thermal electrons can be quantified by $D_{m}$ itself. Therefore, the variance of $D_{m}$ follows
    \begin{equation}\label{eq:vdm}
      V(D_{m}) = \frac{1}{N_{D}}\left({D_{m}} + \sigma_\mathrm{RD}^2\right) + \frac{\sigma_\mathrm{RD}^2}{N_{B}}.
    \end{equation}
    
    \item Master flat $F_{m}$. 
    The raw flat frames $\mathcal{F}$ are obtained by exposing the detector to uniform illumination, either from a dome screen or the twilight sky. Since the pixel counts in flat frames are predominantly determined by the illumination signal, which follows Poisson statistics, the noise contributions from dark current and read noise become negligible. Consequently, the normalized single flat frame can be expressed as $F \simeq (\mathcal{F} - B_{m})/k$, where $k$ is the normalization factor. $F_{m}$ is then constructed through averaging $N_F$ individual flat frames $F_{m} = \sum_{i=1}^{N_{F}}F_{i}/N_{F}$. Its corresponding variance is
    \begin{equation}\label{eq:vfm}
    V(F_{m}) = \frac{1}{N_{F}^2}\sum_{i=1}^{N_{F}}\frac{F_i}{k_{i}}.
    \end{equation}
    Given that raw flat frames are acquired with high signal levels (typically $k >10,000 \times g \, e^{-}$), $V(F_{m})$ is expected to be as small as $\sim 10^{-4}/(gN_{F})$.
    
    \item Background model $S_{m}$. Given the presence of astronomical objects, the conventional approach to model the background involves: (1) dividing the image into rectangular grids, (2) estimating the background level and corresponding root-mean-square (\textsc{rms}) fluctuation within each grid, and (3) applying two-dimensional polynomial interpolation to create smooth background and \textsc{rms} models across the entire image \citep[e.g.,][]{1996A&AS..117..393B}. The \textsc{rms} model can serve as a low-resolution approximation of the weight map, i.e., \(W(T) \approx 1/\textsc{rms}^2\). However, interpolation inherently smooths out local variations, potentially reducing its accuracy. 
    
    \item Variance of raw image $V(R)$. Based on the above analyses, $V(R)$ is given by 
    \begin{equation}\label{eq:vr}
    V(R) = (T+S_m)\,F_{m} + D_{m} + \sigma_\mathrm{RD}^2.
    \end{equation}
\end{itemize}

Substituting equations~(\ref{eq:vbm})-(\ref{eq:vr}) into equation~(\ref{eq:vt1}), $V(T)$ is derived as 

\begin{equation}\label{eq:var}
    V(T) = \frac{T+S_m}{F_{m}} + \frac{s_{D}\,D_{m}}{F_{m}^2} + \frac{s_\mathrm{RD}\,\sigma_\mathrm{RD}^2}{F_{m}^2} + \frac{(T+S_{m})^2}{F_{m}^2\,N_{F}^2}\sum_{i=1}^{N_{F}}\frac{F_i}{k_{i}},
\end{equation}
where $s_{D} = 1 + 1/N_{D}$ and $s_\mathrm{RD} = 1 + 2/N_{B} + 1/N_{D}$.

We have now derived the variance map by accounting for different noise contributions arising from the CCD detectors and standard image processing procedures. Although this represents a complete formulation, an approximate expression can often be employed in practice.
\begin{itemize}
\item For many modern CCD detectors, the dark current is negligible. If we further ignore the read noise contributions in $B_{m}$ and $D_{m}$, and disregard the Poisson noise from astronomical objects, Equation~(\ref{eq:var}) reduces to 
\begin{equation}
V(T) = \frac{S_m}{F_{m}} + \frac{1}{F_{m}^2}\left(\sigma_\mathrm{RD}^2 + \frac{S_{m}^2}{N_{F}^2}\sum_{i=1}^{N_{F}}\frac{F_i}{k_{i}}\right).
\end{equation}

\item Furthermore, we have shown that $V(F_{m})$ is also expected to be negligible. For small background ($S_{m} \ll 100\,\sqrt{gN_{F}} \, e^{-}$), we further have
\begin{equation}
    V(T) = \frac{S_m}{F_{m}} + \frac{\sigma_\mathrm{RD}^2}{F_{m}^2}.
\end{equation}
This simplified form has been used in modern surveys like the Dark Energy Survey \citep{2017PASP..129k4502B}. However, we emphasize that in bright observing conditions (e.g., during lunar illumination), the background-induced term $S_m^2V(F_m)$ may introduce non-negligible noise contribution.
\end{itemize}

Additionally, pixel counts in raw CCD images are conventionally recorded in analog-to-digital units (ADU), and converted to electrons ($e^{-}$) by multiplying the gain $g$. This establishes the variance relationship $V(C) = g^2 \, V(C^\mathrm{ADU})$, where $C$ denotes the electron count, and $C^\mathrm{ADU}$ represents the original ADU measurement. If the units of all quantities are converted back to ADU (while keeping $\sigma_\mathrm{RD}$ in $e^{-}$/pixel), re-writing equation~(\ref{eq:var}) yields 
\begin{equation}\label{eq:var_aud}
    V(T^{\mathrm{ADU}}) = \frac{T^\mathrm{ADU}+S_m^\mathrm{ADU}}{gF_{m}} + \frac{s_{D}\,D_{m}^\mathrm{ADU}}{gF_{m}^2} + \frac{s_\mathrm{RD}\,\sigma_\mathrm{RD}^2}{g^2 F_{m}^2} + \frac{(T^\mathrm{ADU}+S_{m}^\mathrm{ADU})^2}{gF_{m}^2\,N_{F}^2}\sum_{i=1}^{N_{F}}\frac{F_i}{k_{i}^\mathrm{ADU}}.
\end{equation}

In practice, the weight map should incorporate bad pixel masks (e.g., bad columns, cosmic rays, saturated/hot/cold pixels). These problematic pixels are typically assigned zero weight so that they can be excluded from subsequent analyses.

\begin{acknowledgments}
This work is supported by the National Key R\&D Program of China No. 2022YFF0503404, and the Key Laboratory of Survey Science of Yunnan Province with project No. 202449CE340002.

\end{acknowledgments}


\bibliography{sample701}{}
\bibliographystyle{aasjournalv7}



\end{document}